\documentclass[12pt,letter]{article}

\usepackage{mystyle}
\usepackage{mythmstyle_report}

\usepackage{calc}
\usepackage{tikz}
\usetikzlibrary{decorations.markings}
\tikzstyle{vertex}=[circle, draw, inner sep=0pt, minimum size=6pt]
\begin{document}
\title{Fault tolerant supergraphs with automorphisms}

\author{Ashwin~Ganesan%
  \thanks{School of Computer Science and Engineering, Vellore Institute of Technology (VIT University), Vellore, Tamilnadu 632014, India. Email:  
\texttt{ashwin.ganesan@gmail.com}.}
}

\date{}
\vspace{10cm}

\maketitle

\begin{abstract}
\noindent  Given a graph $Y$ on $n$ vertices and a desired level of fault-tolerance $k$, an objective in fault-tolerant system design is to construct a supergraph $X$ on $n + k$ vertices such that the removal of any $k$ nodes from $X$ leaves a graph containing $Y$. In order to reconfigure around faults when they occur, it is also required that any two subsets of $k$ nodes of $X$ are in the same orbit of the action of its automorphism group.  In this paper, we prove that such a supergraph must be the complete graph.  This implies that it is very expensive to have an interconnection network which is $k$-fault-tolerant and which also supports automorphic reconfiguration.  Our work resolves an open problem in the literature. The proof uses a result due to Cameron on $k$-homogeneous groups. 
\end{abstract}


%
%
\section{Introduction}

The interconnection network of a computing system is modeled as a graph $X=(V,E)$ whose vertices correspond to processors and with two vertices being adjacent whenever the corresponding two processors are connected by a direct communication link (cf. \cite{Hayes:1976}, \cite{Leighton:1992}).  In order to execute an algorithm on this computing system, it is required that the architecture $X$ contain a given {\em basic graph} $Y$ as a subgraph.  If some of the nodes of $X$ become faulty, in order to continue operation it is required that the functioning part of the network still contain the basic graph $Y$.  We assume the basic graph $Y$ is nonempty, i.e. it contains at least one edge.  Any notation or terminology on graphs used in this paper which we do not explicitly define here is standard and can be found in \cite{Bollobas:1998}. 

Let $Y$ be a nonempty graph on $n$ vertices.  A graph $X$ is said to be a {\em $k$-fault-tolerant realization of $Y$} if $X$ can be obtained from $Y$ by adding a set of $k$ new vertices (called {\em spare nodes}) and some edges so that the resulting graph $X$ has the property that the removal of any $k$ vertices from $X$ leaves a graph which still contains $Y$ (cf. \cite{Dutt:Hayes:1991}).  In other words, $X$ is a $k$-fault-tolerant realization of $Y$ if $X$ has exactly $n+k$ vertices and $X-W$ contains a subgraph isomorphic to $Y$ for each $k$-subset $W \subseteq V(X)$.  In this case, if any $k$ nodes of $X$ become faulty, the network corresponding to the nonfaulty nodes of $X$ contains the architecture $Y$ and hence can continue to operate.  In this sense, the architecture $X$ can tolerate up to $k$ node failures. 

A graph $X$ is a \emph{supergraph} of $Y$ if it is possible to add vertices and edges to $Y$ to obtain $X$, i.e. if $V(Y) \subseteq V(X)$ and $E(Y) \subseteq E(X)$.  A graph $X$ is an \emph{edge-supergraph} of $Y$ if $V(X)=V(Y)$ and $E(Y) \subseteq E(X)$. In this paper, we consider the method of constructing supergraphs $X$ of a given graph $Y$ such that for all $k$-subsets $F \subseteq V(X)$, $X-F$ contains the subgraph $Y$.  This type of design method is called \emph{global sparing} because the $k$ spare nodes added to $Y$ are associated with all of $V(Y)$. In local sparing, we would partition $V(Y)$ into $t$ subsets $V_1,\ldots,V_t$ (for some $t \ge 2$) and associate $t$ sets of spare nodes to the $t$ subsets $V_i$, respectively, such that $k_i$ spare nodes are associated with $V_i$ $(i=1,\ldots,k)$ and $k_1+\ldots+k_t = k$.  Local sparing simplifies the design and reconfiguration process, while global sparing achieves $k$-fault-tolerance with fewer processors.

Many authors have investigated the use of algebraic methods in interconnection networks; see the surveys \cite{Heydemann:1997}, \cite{Ganesan:PCWAAC:2016} and the references therein.  The present paper investigates an open problem posed in \cite{Dutt:Hayes:1991} on fault-tolerant supergraphs whose automorphism group satisfies certain properties.  Another research area at the interface of graph automorphisms and interconnection networks is the study of the structure of interconnection network topologies; for example, several authors have investigated the automorphism group of graphs that arise as the topology of interconnection networks \cite{Deng:Zhang:2012} \cite{Feng:2006} \cite{Ganesan:DM:2013}   \cite{Ganesan:DMGT} \cite{Ganesan:JACO}   \cite{Wan:Zhang:2009}.  More recently, researchers have questioned the practicability and advantages of interconnection networks with large automorphism groups; for example, data center interconnection networks are not hyperbolic \cite{Coudert:Ducoffe:2016}. 

For basic definitions of terms such as groups, symmetric groups, and homomorphisms, we refer the reader to \cite[Chapter 1]{Dixon:Mortimer:1996}, \cite{Cameron:1999}.  Let $G$ be a group and let $\Omega$ be a nonempty set.  Suppose the map $\mu: \Omega \times G \rightarrow \Omega, (\alpha,x) \mapsto \alpha^x$ satisfies the following two conditions: (i) $(\alpha^x)^y = \alpha^{xy}$, for all $\alpha \in \Omega$ and all $x,y \in G$, and (ii) $\alpha^1 = \alpha$ for all $\alpha \in \Omega$, where $1$ denotes the identity element of the group $G$.  Then, we say that this map defines an \emph{action} of $G$ on $\Omega$ and that $G$ \emph{acts} on $\Omega$.  This action naturally induces a homomorphism from $G$ into the symmetric group $\Sym(\Omega)$, and so each element of $G$ induces a permutation of $\Omega$. Conversely, every homomorphism from $G$ into $\Sym(\Omega)$ induces an action of $G$ on $\Omega$.  The orbit of a point $\alpha \in \Omega$ under this action is the set $\alpha^G := \{\alpha^x: x \in G\}$. Thus, the action of $G$ on $\Omega$ partitions $\Omega$ into orbits. The action of $G$ on $\Omega$ is  \emph{transitive} if for all $\alpha, \beta \in \Omega$, there exists a $g \in G$ such that $\alpha^g = \beta$; equivalently, $G$ acts transitively on $\Omega$ if the action of $G$ on $\Omega$ has a single orbit. 

Suppose $G$ acts on $\Omega$.  The action of $G$ on $\Omega$ induces an action of $G$ on the set of all subsets of $\Omega$ by the rule $\Gamma^x := \{\gamma^x: \gamma \in \Gamma\}$, for all $\Gamma \subseteq \Omega$.  It is clear that the set $\Omega^{\{k\}}$ of all $k$-subsets of $\Omega$ is $G$-invariant, i.e. $(\Omega^{\{k\}})^x = \Omega^{\{k\}}$ for all $x \in G$.   The group $G$ is said to be {\emph{$k$-homogeneous} if $G$ acts on $\Omega^{\{k\}}$ transitively.   (What we call $k$-homogeneous in this paper is referred to as $k$-subtransitive in \cite{Dutt:Hayes:1991}.)  

A \emph{$k$-tuple} of distinct elements from $\Omega$ is an ordered subset of $k$ distinct elements from $\Omega$. For example, if $\{\delta_1,\ldots,\delta_k\}$ is a $k$-subset of $\Omega$, then $(\delta_1,\ldots,\delta_k)$ is a $k$-tuple of distinct elements from $\Omega$.  Let $\Omega^{(k)}$ denote the set of all $k$-tuples of distinct elements from $\Omega$.  We say that $G$ is \emph{$k$-transitive} if $G$ acts transitively on $\Omega^{(k)}$. Thus, the action of $G$ on $\Omega$ is $k$-transitive iff for every two $k$-tuples $(\alpha_1,\ldots,\alpha_k)$, $(\beta_1,\ldots,\beta_k)$ of distinct elements from $\Omega$, there exists a $g \in G$ such that $\alpha_i^g=\beta_i$ $(i=1,\ldots,k$).   For further details on group actions, we refer the reader to \cite[Chapter 1]{Dixon:Mortimer:1996}; an introduction to multiply transitive groups and $k$-homogeneous groups can be found in 
\cite[Chapter II]{Wielandt:1964} and \cite[Sections 2.1 and 9.4]{Dixon:Mortimer:1996}.

Let $X=(V,E)$ be a simple, undirected graph.  Let $\Sym(V)$ denote the full symmetric group acting on the vertex set $V=V(X)$.  Then, $\Sym(V)$ acts naturally on the set $V^{\{2\}}$ of all $2$-subsets of $V$ by the following rule: for all $x \in \Sym(V)$ and for all $ \{u, v\}  \in V^{\{2\}}$, $\{u,v\}^x := \{u^x, v^x\}$.  An \emph{automorphism} of the graph $X=(V,E)$ is a permutation $g \in \Sym(V)$ which preserves adjacency and nonadjacency. In other words, $g \in \Sym(V)$ is an automorphism of $X$ if $\{x,y\} \in E$ iff $\{x,y\}^g \in E$.  The set of all automorphisms of $X$ forms a permutation group, called the \emph{automorphism group of $X$}, denoted by $\Aut(X)$.  Thus,  $\Aut(X) := \{g \in \Sym(V): E^g = E\}$.   A graph $X$ is said to be \emph{vertex-transitive} if its automorphism group $\Aut(X)$ acts transitively on the vertex set $V(X)$.  For an introduction to automorphisms of graphs, the reader is referred to  \cite{Godsil:Royle:2001}.

Having stated our terminology on group actions and automorphisms of graphs, we can now describe an approach for restructuring around faults in an interconnection network.  This approach, called automorphic reconfiguration, was introduced in \cite{Dutt:Hayes:1991}, and refers to a specific type of reconfiguration using automorphisms in which the $k$ spare nodes are directly mapped to the set of $k$ faulty nodes.  Automorphic reconfiguration, as defined in \cite{Dutt:Hayes:1991}, is an impractical way of designing and reconfiguring graphs (and in particular, multiprocessor networks), and was a definition given just for theoretical purposes to finally lead to the type of $k$-fault-tolerant supergraphs designed in \cite{Dutt:Hayes:1991} that are not complete graphs.

In order to achieve so-called {\em automorphic reconfiguration} (cf. \cite[p.253]{Dutt:Hayes:1991}), it is required that, when $k$ or fewer nodes of the interconnection network become faulty, there exists an automorphism of the graph $X$ that maps the spare nodes to the faulty nodes.  During this reconfiguration process, the faulty nodes are relabeled as spare nodes, and the nonfaulty nodes are relabeled as nodes of $Y$ and contain a subgraph isomorphic to $Y$.  In graph-theoretic terms, the interconnection topology $X$ must satisfy the property that if $A$ and $B$ are any two $k$-subsets of $V(X)$, then there is an automorphism of $X$ that maps $A$ to $B$.   Equivalently, the interconnection network topology $X$ must satisfy the property that its automorphism group $\Aut(X)$ is $k$-homogeneous.  


Thus, our objective is the following: given a basic graph $Y$ and a desired level of fault-tolerance $k$, construct a graph $X$ such that $X$ is a $k$-fault-tolerant realization of $Y$ and such that $\Aut(X)$ is $k$-homogeneous.  In other words, given a basic graph $Y$, we add $k$ spare nodes $S$ to $V(Y)$ and edges to get a supergraph $X$ such that for any $k$-subset $F \subseteq V(X)$ of faulty nodes, there exists an automorphism $g \in \Aut(X)$ such that $S^g = F$ and such that the set of nonfaulty nodes contains the subgraph $Y$.  Dutt and Hayes settled this problem for the case $k=2$  by proving the following result:  

\begin{Theorem} \cite[Theorem 2]{Dutt:Hayes:1991} \label{thm:DH:k:eq:2}
If $Y$ is a nonempty graph on $n$ vertices, $X$ is a 2-fault-tolerant realization of $Y$ and  $\Aut(X)$ is 2-homogeneous, then $X$ is the complete graph $K_{n+2}$.   
\end{Theorem}

Dutt and Hayes (cf. \cite[p. 253]{Dutt:Hayes:1991}) posed the problem of generalizing the $k=2$ result of Theorem~\ref{thm:DH:k:eq:2} to arbitrary $k$.  In this paper, we resolve this open problem (cf. Theorem~\ref{thm:k:DuttHayes:conjecture} below).  

The following is the main result of this paper:

\begin{Theorem}
 \label{thm:k:DuttHayes:conjecture}
 Let $k \ge 2$.  If $Y$ is a nonempty graph on $n$ vertices, $X$ is a $k$-fault-tolerant realization of $Y$ and $\Aut(X)$ is $k$-homogeneous, then $X$ is the complete graph $K_{n + k}$. 
\end{Theorem}

We point out that if $Y$ has $n$ vertices and $X$ is a $k$-fault-tolerant realization of $Y$, then $X$ has exactly $n + k$ vertices.  The condition that $X$ have exactly $n+k$ vertices can be relaxed; see Corollary~\ref{cor:X:numVertices} and the remarks preceding it.  One way to state the more general version of our main result is as follows: If $Y$ is a nonempty graph, and $X$ is such that $\Aut(X)$ is $k$-homogeneous for some $k \ge 2$ and the removal of any $k$ nodes from $X$ leaves a graph containing $Y$, then $X$ is a complete graph.  We say ``a'' complete graph, rather than ``the'' complete graph because the order of $X$ can be arbitrary.  In most of this paper, we follow the statement of the open problem in \cite[p. 253]{Dutt:Hayes:1991} and so assume that $X$ has exactly $n+k$ vertices.
\section{Preliminaries}
%
%
%
%
%
%
%

We recall some results on permutation groups.

\begin{Lemma}\label{lem:ktrans:implies:khom}
 Suppose $G$ acts on $\Omega$. If $G$ is $k$-transitive, then $G$ is $k$-homogeneous.
\end{Lemma}

\noindent \emph{Proof:}  
Let $A = \{a_1,\ldots, a_k\}$ and $B = \{b_1,\ldots, b_k\}$ be $k$-subsets of $\Omega$.  
To show $G$ is $k$-homogeneous, it suffices to show there exists $g \in G$ such that $A^g = B$.   By $k$-transitivity of $G$, there exists $g \in G$ such that $(a_1, \ldots, a_k)^g = (b_1, \ldots, b_k)$, i.e. there exists $g \in G$ such that $a_i^g = b_i$ $(i=1,\ldots,k)$. Hence, $A^g = \{a_1^g, \ldots, a_k^g\} = B$.
\qed

\begin{Lemma} \label{lem:khom:nminuskhom} \cite[p. 35]{Dixon:Mortimer:1996}
 Suppose $G$ acts on $\Omega$ and $|\Omega|=n$.  Then, $G$ is $k$-homogeneous iff $G$ is $(n-k)$-homogeneous.
\end{Lemma}

\noindent \emph{Proof:} Suppose $G$ is $k$-homogeneous.  To show $G$ is $(n-k)$-homogeneous, let $A, B \in \Omega^{\{n-k\}}$.  It suffices to show there exists a $g \in G$ such that $A^g = B$.  Let $A' := \Omega - A$, $B' := \Omega - B$.  Then, $A', B' \in \Omega^{\{k\}}$.  By hypothesis, there exists a $g \in G$ such that $(A')^g = B'$.  Since $g$ acts on $\Omega$, $g$ takes the complement of $A'$ to the complement of $B'$, i.e. $A^g = B$. The converse is proved in a similar manner. 
\qed

\begin{Lemma} \label{lem:ktrans:implies:kminus1trans}
Suppose $G$ acts on $\Omega$, $|\Omega| = n$ and $2 \le k \le n$. If $G$ is $k$-transitive, then $G$ is $(k-1)$-transitive.
\end{Lemma}

\noindent \emph{Proof:}
Let $\Delta = \{\delta_1,\ldots,\delta_{k-1}\}$ and $\Gamma = \{\gamma_1,\ldots,\gamma_{k-1}\}$ be $k$-subsets of $\Omega$. Let $\alpha \in \Omega - \Delta$ and $\beta \in \Omega - \Gamma$.  By $k$-transitivity of $G$, there exists a $g \in G$ such that $(\delta_1,\ldots,\delta_{k-1},\alpha) ^ g$ $=(\gamma_1,\ldots,\gamma_{k-1},\beta)$.   Hence, there exists a $g \in G$ which takes the tuple $(\delta_1,\ldots,\delta_{k-1})$ to the tuple $(\gamma_1,\ldots,\gamma_{k-1})$. 
\qed

If $|\Omega| = n$, we write $S_n$ for the full symmetric group $\Sym(\Omega)$.  The group of $n!$ permutations in $S_n$ acts naturally on the set $[n] := \{1,\ldots,n\}$, and it is clear that for any two $k$-tuples $\alpha = (\alpha_1,\ldots,\alpha_k)$, $\beta = (\beta_1,\ldots,\beta_k)$ $(k \le n)$, there exists a $g \in S_n$ such that $\alpha^g = \beta$.  Hence, $S_n$ is $k$-transitive for every $k \in [n]$. By Lemma~\ref{lem:ktrans:implies:khom}, $S_n$ is also $k$-homogeneous for every $k \in [n]$.  The next result shows some formal distinction between $k$-homogeneity and $k$-transitivity by exhibiting a permutation group (subgroup of $S_n$) which is $2$-homogeneous and not $2$-transitive.

\begin{Lemma} \cite[p. 286]{Dixon:Mortimer:1996}
Consider the permutations $x = (12\ldots7), y = (235)(476)$ in $S_7$.  Let $G:=\langle x, y \rangle$ be the permutation group in $S_7$ generated by $x$ and $y$. Then, $G$ is not $2$-transitive and $G$ is $2$-homogeneous. 
\end{Lemma}

\noindent \emph{Proof:}  Observe that $xy = yx^2$ and so $|G| = 21$.  If the action of a group on $[7]$ is $2$-transitive, then the group must contain at least $42$ elements. Because the order of $G$ is $21$, $G$ is not $2$-transitive. 

To show $G$ is $2$-homogeneous, suppose $\{\alpha, \beta\} \in [7]^{\{2\}}$.  Then, there exists $z \in \langle x \rangle$ such that $\{\alpha, \beta\}^z = \{1, \gamma\}$ for some $\gamma$.  If $\gamma \in \{2,3,5\}$, then $\{1,\gamma\}^{y^i} = \{1,2\}$ for some $i$.  If $\gamma \in \{4,7,6\}$, then $\{1, \gamma\}^{y^j} = \{1,7\}$ for some $j$, and we know $\{1,7\}^x = \{1,2\}$.  Thus, there exists $g \in G$ such that $\{\alpha,\beta\}^g = \{1,2\}$.  Hence, the action of $G$ on $[7]^{\{2\}}$ has a single orbit.
\qed

Recall that if a group is $k$-transitive, then it is also $(k-1)$-transitive.  In general, a group which is $k$-homogeneous is not necessarily $(k-1)$-homogeneous.  It will follow from our main result that if a permutation group $G$ arises as the automorphism group of a graph, then $k$-homogeneity of $G$ does imply $(k-1)$-homogeneity of $G$.

The proof of the main result uses the following result due to Cameron:

\begin{Theorem} \cite[Theorem 2.2]{Cameron:1976} \cite[Theorem 9.4A]{Dixon:Mortimer:1996} \label{thm:cameron:1976}
 Let $G$ be a permutation group acting on a set $\Omega$.  Let $m, k$ be integers with $0 \le m \le k$ and $m+k \le |\Omega|$.  Then, $G$ has at least as many orbits in $\Omega^{\{k\}}$ as it has in $\Omega^{\{m\}}$.
\end{Theorem}

\section{Main results}

We first extend Theorem~\ref{thm:DH:k:eq:2} from the $2$-homogeneous case to the $3$-homogeneous case.  This result is actually a special case of the main result (Theorem~\ref{thm:k:DuttHayes:conjecture}). We now give an elementary proof for the $3$-homogeneous case which does not use the theory of permutation groups. In fact, the condition that $\Aut(X)$ be $3$-homogeneous can be replaced by the weaker condition that every subset of $3$ vertices of the graph induces the same subgraph.

\begin{Theorem}\label{thm:3hom:complete}
If $X$ is a graph on $5$ or more vertices containing at least one edge and if every subset of $3$ vertices of $X$ induces the same subgraph, then $X$ is a complete graph. 
\end{Theorem}

\noindent \emph{Proof:} Let $A=\{a,b,c\} \subseteq V(X)$ and let $X'$ denote the induced subgraph $X[A]$.  Because $\Aut(X)$ is $3$-homogeneous, every $3$-subset of $V(X)$ induces the same subgraph $X'$.  By hypothesis, $X'$ contains at least one edge.  We consider three cases for the structure of $X'$.  For the first case, suppose $X'$ is a $3$-clique. Then, $X$ is the complete graph because every subset of $3$ vertices of $X$ induces a $3$-clique. For the second case, suppose $X'$ is isomorphic to $K_{1,2}$.  Without loss of generality, suppose $ab, bc \in E(X), ac \notin E(X)$.  Let $x,w \in V(X)-A$.  Then, $\{a,c,w\}$ and $\{a,c,x\}$ each induce a $K_{1,2}$. Since $a$ and $c$ are nonadjacent, it must be that $ax, cx, aw, cw \in E(X)$. The subgraph induced by $\{b, x, w\}$ must also be a $K_{1,2}$ and hence contains at least one edge.  The endpoints of this edge along with vertex $a$ induce a $K_3$, a contradiction.  Hence, the second case is impossible.  

For the third case, suppose the induced subgraph $X'$ is isomorphic to the disjoint union of $K_2$ and $K_1$.  Then, the complement graph $\overline{X'}$ is isomorphic to $K_{1,2}$.  Since the automorphism group of a graph and of its complement are equal, $\Aut(\overline{X})$ is also $3$-homogeneous, and by the second case above, $\overline{X}$ is the complete graph. But this implies $X$ is the empty graph, contradicting the fact that its induced subgraph $X'$ contains an edge.  Hence, this third case is also impossible.
\qed

The lower bound of $5$ in Theorem~\ref{thm:3hom:complete} is tight since the automorphism group of $C_4$ is $3$-homogeneous and $C_4$ is not a clique. 

\bigskip We now prove the main result.

\noindent \emph{Proof of Theorem~\ref{thm:k:DuttHayes:conjecture}:}
The case $k=2$ is addressed in Theorem~\ref{thm:DH:k:eq:2}, so assume $k \ge 3$.  Let $Y$ be a graph on $n$ vertices. Here, $n \ge 2$ since $Y$ contains at least one edge.   Note that $X$ is a graph on $n+k$ vertices.  Let $G=\Aut(X)$ and let $\Omega = V(X)$.  By hypothesis, the action of $G$ on the set $\Omega$ is $k$-homogeneous.  Thus, the number of orbits of $G$ on $\Omega^{\{k\}}$ is 1.  Since $2 \le n$, $2+k \le n+k = |\Omega|$.  Also, $2 \le k$. Hence, by Theorem~\ref{thm:cameron:1976}, the number of orbits of $G$ on $\Omega^{\{2\}}$ is also 1.  Equivalently, the action of $G$  on $\Omega$ is 2-homogeneous. 

Let $\{u,v\}$ be an edge in $Y$.  Then $\{u,v\}$ is an edge in $X$.  Let $a$ and $b$ be distinct vertices of $X$.  Because $G$ is $2$-homogeneous, there is an element $g \in G$ that maps $\{u,v\}$ to $\{a,b\}$.  The automorphism $g$ preserves adjacency, whence $\{a,b\}$ is an edge of $X$.  This proves that any two distinct vertices in $X$ are adjacent, i.e. $X$ is the complete graph $K_{n+k}$.
\qed

In the proof above, we essentially showed the following:

\begin{Corollary}
 Let $k \ge 2$.  Let $X$ be a nonempty graph on $k+2$ vertices.  If $\Aut(X)$ is $k$-homogeneous, then $X$ is the complete graph $K_{k+2}$.
\end{Corollary}

As stated, Theorem~\ref{thm:k:DuttHayes:conjecture}  assumes that the order $|V(X)|$ of $X$ is exactly equal to $n+k$.  The proof of the theorem goes through even if this condition is relaxed to $|V(X)| \ge n+k$:

\begin{Corollary} \label{cor:X:numVertices}
 Let $X$ be a nonempty graph.  If $\Aut(X)$ is $k$-homogeneous for some $2 \le k \le |V(X)|-2$, then $\Aut(X)$ is $2$-homogeneous and $X$ is a complete graph. 
\end{Corollary}

We resolve some further open questions in the literature. In \cite[p. 252]{Dutt:Hayes:1991}, it is stated that there is likely no characterization of $k$-homogeneous automorphism groups of graphs in the literature.  In \cite[p. 252]{Dutt:Hayes:1991}, it is mentioned that $k$-homogeneity of $\Aut(X)$ could be a ``significantly weaker restriction'' than $k$-transitivity of $\Aut(X)$. Also, in \cite[p.253]{Dutt:Hayes:1991}, it is mentioned that $k$-homogeneity of $\Aut(X)$ does not necessarily imply $i$-homogeneity of $\Aut(X)$ for $2 \le i \le k-1$.  All these questions are resolved by the following immediate consequence of our main result Theorem~\ref{thm:k:DuttHayes:conjecture}:

\begin{Corollary}
 Let $X$ be a graph on $4$ or more vertices. Suppose $2 \le k \le |V(X)|-2$.  If $\Aut(X)$ is $k$-homogeneous, then $\Aut(X)$ is $i$-homogeneous for $i=1,2,\ldots,k-1$ and $\Aut(X)$ is the full symmetric group.  In particular, $k$-homogeneity of $\Aut(X)$ is equivalent to $k$-transitivity of $\Aut(X)$. 
\end{Corollary}

The bounds in $2 \le k \le |V(X)|-2$ are tight - there are many families of vertex-transitive graphs  for which $\Aut(X)$ is $k$-homogeneous for $k=1$ (and hence also for $k=|V(X)|-1$ by Lemma~\ref{lem:khom:nminuskhom}, and obviously for $k=|V(X)|$) and such that $\Aut(X)$ is not $i$-homogeneous for $i=2,\ldots,|V(X)|-2$.  For example, take $X$ to be a cycle graph.

Theorem~\ref{thm:k:DuttHayes:conjecture} makes two assumptions: that $X$ is a $k$-fault-tolerant realization of $Y$ and that $\Aut(X)$ is $k$-homogeneous.  The conclusion of the theorem follows mainly from the second assumption and only a weak consequence of the first assumption is used.  To illustrate, we consider the following example of a graph $X$ which is a $2$-fault-tolerant realization of $Y=Q_3$, the $3$-dimensional cube.  As this example shows, the first assumption alone is not sufficient to ensure that $X$ is the complete graph.

\begin{Theorem}\label{thm:3cube:2FT:notkhom}
 Let $Y$ be the $3$-dimensional hypercube $Q_3$.  Let $X$ be the graph obtained by adding to $Y$ two spare nodes $x_1$ and $x_2$ and joining each $x_i$ ($i=1,2$) to each vertex of $Y$.  Then, $X$ is a 2-fault-tolerant realization of $Y$, but $\Aut(X)$ is not $k$-homogeneous for $k = 1,2,\ldots,|V(X)|-1$.
\end{Theorem}

\noindent \emph{Proof:}
Observe that if any two diametrically opposite vertices of the $3$-cube $Q_3$ are removed, the resulting graph is a $6$-cycle graph; see Figure\ref{fig:3cube}(a).  Adding two new vertices to this $6$-cycle graph and joining these two vertices to each vertex of the $6$-cycle gives an edge-supergraph of $Q_3$.  Hence, if a graph contains a $6$-cycle, then adding two new vertices $x_1,x_2$ to the graph and joining each $x_i$ to each vertex of the graph gives an edge-supergraph of $Q_3$. 

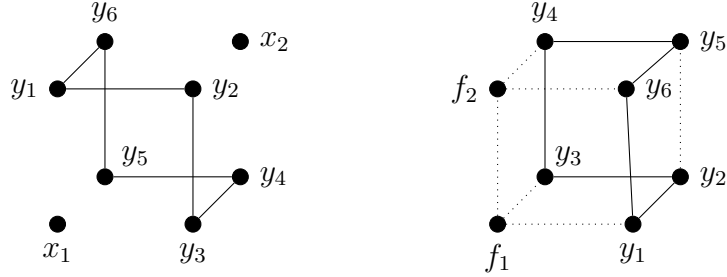
\begin{figure}
\begin{centering}
 \begin{tikzpicture}[scale=0.9,auto]
 \pgfmathsetmacro{\phi}{0.7}
 \pgfmathsetmacro{\delx}{6.5} 
 
  \vertex[fill] (v000) at (0,0) [label=below:{$x_1$}] {};
  \vertex[fill] (v001) at (2,0) [label=below:{$y_3$}] {};
  \vertex[fill] (v100) at (0,2) [label=left:{$y_1$}] {};
  \vertex[fill] (v101) at (2,2) [label=right:{$y_2$}] {};

  \vertex[fill] (v010) at (0+\phi,0+\phi) {};
  \node at (0.1+\phi+0.35,0+\phi+0.3) {$y_5$};
  \vertex[fill] (v011) at (2+\phi,0+\phi) [label=right:{$y_4$}] {};
  \vertex[fill] (v110) at (0+\phi,2+\phi) [label=above:{$y_6$}] {};
  \vertex[fill] (v111) at (2+\phi,2+\phi) [label=right:{$x_2$}] {};

  \draw (v001) -- (v101) -- (v100);
  \draw (v010) -- (v011);
  \draw (v001) -- (v011);  
  \draw (v100) -- (v110) -- (v010);
  
  \vertex[fill] (w000) at (0+\delx,0) [label=below:{$f_1$}] {};
  \vertex[fill] (w001) at (2+\delx,0) [label=below:{$y_1$}] {};
  \vertex[fill] (w100) at (0+\delx,2) [label=left:{$f_2$}] {};
  \vertex[fill] (w101) at (1.9+\delx,2) [label=right:{$y_6$}] {};

  \vertex[fill] (w010) at (0+\phi+\delx,0+\phi) {};
  \node at (0+\phi+0.35+\delx,0+\phi+0.3) {$y_3$};
  \vertex[fill] (w011) at (2+\phi+\delx,0+\phi) [label=right:{$y_2$}] {};
  \vertex[fill] (w110) at (0+\phi+\delx,2+\phi) [label=above:{$y_4$}] {};
  \vertex[fill] (w111) at (2+\phi+\delx,2+\phi) [label=right:{$y_5$}] {};

  \draw (w001) -- (w101);
  \draw[dotted] (w000) -- (w001);
  \draw[dotted] (w100) -- (w101);
  \draw[dotted] (w000) -- (w100);
  \draw (w010) -- (w011);
  \draw (w111) -- (w110) -- (w010);
  \draw[dotted] (w011) -- (w111);
  \draw (w001) -- (w011);
  \draw[dotted] (w000) -- (w010);
  \draw[dotted] (w100) -- (w110);
  \draw (w101) -- (w111);
  
\end{tikzpicture}
\caption{(a) Removing diametrically opposite vertices from a cube $Q_3$ gives a $6$-cycle $y_1 y_2 \ldots y_6$. (b) Removing two adjacent vertices gives a graph containing a $6$-cycle $y_1 y_2 \ldots y_6$.}
\label{fig:3cube}
\end{centering}
\end{figure}

To prove that $X$ is a 2-fault-tolerant realization of $Y=Q_3$, let $F=\{f_1,f_2\}$ be a set of two faulty nodes of $X$. We need to show that $X-F$ contains a subgraph isomorphic to $Q_3$. This is clear if $\{f_1,f_2\} = \{x_1,x_2\}$ and also if exactly one of the faulty nodes is in $Y$ because this node can be replaced by the non-faulty spare node.  So suppose now that both faulty nodes $f_1$ and $f_2$ are in $Y$. If $f_1$ and $f_2$ are nonadjacent in $Y$, then they can be replaced by the spare nodes $x_1$ and $x_2$, respectively, and $Y-F$ contains a subgraph isomorphic to $Q_3$.  Finally, suppose $f_1$ and $f_2$ are adjacent nodes of $Y$.  Then $Y-F$ contains a $6$-cycle (see Figure~\ref{fig:3cube}(b)), and by the argument in the previous paragraph, $X-F$ contains a subgraph isomorphic to $Q_3$.

In the graph $X$, the degree of vertex $x_i$ ($i=1,2$) is $8$, and the degree of each of the remaining vertices is $4$. Hence, $X$ is not vertex-transitive (i.e. $\Aut(X)$ is not $1$-homogeneous).  By Lemma~\ref{lem:khom:nminuskhom}, $\Aut(X)$ is not $(|V(X)|-1)$-homogeneous. If $\Aut(X)$ is $k$-homogeneous for some $2 \le k \le |V(X)|-2$, then by Theorem~\ref{thm:k:DuttHayes:conjecture} $X$ is the complete graph, a contradiction.  Hence, $\Aut(X)$ is not $k$-homogeneous if $k \in \{ 1,2,\ldots, |V(X)|-1\}$. 
\qed

The paper \cite{Dutt:Hayes:1991} designed an iterative reconfiguration technique which after $k$ faults occur uses $k$ different automorphisms in a repeated manner to obtain a fault-free graph isomorphic to the basic graph $Y$ from its $k$-fault-tolerant supergraph $X$ that does not require $X$ to be a complete graph, and in fact is quite efficient in the additional edges needed in $X$ with respect to $Y$ (cf. \cite[Theorems 5 and 6]{Dutt:Hayes:1991}) as well as in the switch-based implementation of $X$.  Both Theorem~\ref{thm:3cube:2FT:notkhom} and the technique in \cite{Dutt:Hayes:1991} are evidence that $k$-fault-tolerance of $X$ is not a sufficient condition for $X$ to be complete, and further, in the case of the technique in \cite{Dutt:Hayes:1991} the use of automorphisms to obtain $k$-fault-tolerance does not require $X$ to be complete (and thus does not require $\Aut(X)$ to be $k$-homogeneous).

Theorem~\ref{thm:DH:k:eq:2} and Theorem~\ref{thm:k:DuttHayes:conjecture} imply that it is very expensive to have an interconnection network which is $k$-fault-tolerant and which also supports automorphic reconfiguration (i.e. for $\Aut(X)$ of the $k$-fault-tolerant graph $X$ to also be $k$-homogeneous) because such an interconnection network must be the complete graph.

\section*{Acknowledgements}

Thanks are due to the anonymous reviewers for helpful suggestions.

{
\bibliographystyle{plain}
\bibliography{refsaut}

\begin{thebibliography}{10}

\bibitem{Bollobas:1998}
B.~Bollob\'as.
\newblock {\em Modern Graph Theory}.
\newblock Graduate Texts in Mathematics vol. 184, Springer, New York, 1998.

\bibitem{Cameron:1976}
P.~J. Cameron.
\newblock Transitivity of permutation groups on unordered sets.
\newblock {\em Mathematische Zeitschrift}, 148:127--139, 1976.

\bibitem{Cameron:1999}
P.~J. Cameron.
\newblock {\em Permutation Groups}.
\newblock London Mathematical Society Student Texts 45, Cambridge University
  Press, 1999.

\bibitem{Coudert:Ducoffe:2016}
D.~Coudert and G.~Ducoffe.
\newblock Data center interconnection networks are not hyperbolic.
\newblock {\em Theoretical Computer Science}, 639:72--90, 2016.

\bibitem{Deng:Zhang:2012}
Y.-P. Deng and X.-D. Zhang.
\newblock Automorphism groups of the pancake graphs.
\newblock {\em Information Processing Letters}, 112:264--266, 2012.

\bibitem{Dixon:Mortimer:1996}
J.~D. Dixon and B.~Mortimer.
\newblock {\em Permutation Groups}.
\newblock Graduate Texts in Mathematics vol. 163, Springer, 1996.

\bibitem{Dutt:Hayes:1991}
S.~Dutt and J.~P. Hayes.
\newblock Designing fault-tolerant systems using graph automorphisms.
\newblock {\em Journal of Parallel and Distributed Computing}, 12:249--268,
  1991.

\bibitem{Feng:2006}
Y-Q. Feng.
\newblock Automorphism groups of {C}ayley graphs on symmetric groups with
  generating transposition sets.
\newblock {\em Journal of Combinatorial Theory Series B}, 96:67--72, 2006.

\bibitem{Ganesan:DM:2013}
A.~Ganesan.
\newblock Automorphism groups of {C}ayley graphs generated by connected
  transposition sets.
\newblock {\em Discrete Mathematics}, 313:2482--2485, 2013.

\bibitem{Ganesan:PCWAAC:2016}
A.~Ganesan.
\newblock Cayley graphs and symmetric interconnection networks.
\newblock In {\em Proceedings of the Pre-Conference Workshop on Algebraic and
  Applied Combinatorics (31st Annual Conference of the Ramanujan Mathematical
  Society), Trichy, Tamilnadu, India}, pages 118--170, 2016.

\bibitem{Ganesan:DMGT}
A.~Ganesan.
\newblock Edge-transitivity of {C}ayley graphs generated by transpositions.
\newblock {\em Discussiones Mathematicae Graph Theory}, 36:1035--1042, 2016.

\bibitem{Ganesan:JACO}
A.~Ganesan.
\newblock Automorphism group of the complete transposition graph.
\newblock {\em Journal of Algebraic Combinatorics}, 42(3):793--801, November
  2015.

\bibitem{Godsil:Royle:2001}
C.~Godsil and G.~Royle.
\newblock {\em Algebraic Graph Theory}.
\newblock Graduate Texts in Mathematics vol. 207, Springer, New York, 2001.

\bibitem{Hayes:1976}
J.~P. Hayes.
\newblock A graph model for fault-tolerant computing systems.
\newblock {\em IEEE Transactions on Computers}, C-25(9):875--884, 1976.

\bibitem{Heydemann:1997}
M.~C. Heydemann.
\newblock Cayley graphs and interconnection networks.
\newblock In {\em Graph symmetry: algebraic methods and applications (Editors:
  Hahn and Sabidussi)}, pages 167--226. Kluwer Academic Publishers, Dordrecht,
  1997.

\bibitem{Leighton:1992}
F.~T. Leighton.
\newblock {\em Introduction to parallel algorithms and architectures: arrays,
  trees, hypercubes}.
\newblock Morgan Kaufmann Publishers, 1992.

\bibitem{Wan:Zhang:2009}
M.~Wan and Z.~Zhang.
\newblock A kind of conditional vertex-connectivity of star graphs.
\newblock {\em Applied Mathematics Letters}, 22:264--267, 2009.

\bibitem{Wielandt:1964}
H.~Wielandt.
\newblock {\em Finite Permutation Groups}.
\newblock Academic Press, 1964.

\end{thebibliography}
}
\end{document}